\documentclass[rnote, usenatbib, usepslatex, traditabstract]{aa}
\usepackage{txfonts}
\usepackage{epsfig}
\usepackage{graphicx}
\usepackage{natbib}

\title{Corrections of Sun-as-a-star p-mode frequencies for effects of the solar cycle}

\author{A.-M. Broomhall\inst{1}\thanks{amb@bison.ph.bham.ac.uk} \and W.~J.
Chaplin\inst{1} \and
Y. Elsworth\inst{1} \and S.~T. Fletcher\inst{2} \and R. New\inst{2}}
\institute{School of Physics and Astronomy, University of
Birmingham, Edgbaston, Birmingham B15 2TT \and Faculty of Arts,
Computing, Engineering and Sciences, Sheffield Hallam University,
Sheffield S1 1WB}
\date{}
\begin{document}

\abstract {Solar p-mode frequencies vary with solar activity. It is
important to take this into account when comparing the frequencies
observed from epochs that span different regions of the solar cycle.
We present details of how to correct observed p-mode frequencies for
the effects of the solar cycle. We describe three types of
correction. The first allows mode frequencies to be corrected to a
nominal activity level, such as the canonical quiet-Sun level. The
second accounts for the effect on the observed mode frequencies,
powers, and damping rates of the continually varying solar cycle and
is pertinent to frequencies obtained from very long data sets. The
third corrects for Sun-as-a-star observations not seeing all
components of the modes. Suitable combinations of the three
correction procedures allow the frequencies obtained from different
sets of data to be compared and enable activity-independent
inversions of the solar interior. As an example of how to apply the
corrections we describe those used to produce a set of definitive
Sun-as-a-star frequencies.}


\titlerunning{Solar cycle corrections of Sun-as-a-star
frequencies}\authorrunning{Broomhall et al.} \maketitle
\section{Introduction}

The Birmingham Solar Oscillations Network (BiSON) observes
Sun-as-a-star Doppler velocities. BiSON data can, therefore, be used
to determine the frequencies of low-degree (low-$l$) oscillations.
Low-$l$ oscillations are particularly important as they are the only
p-modes that penetrate the solar core.

Accurate and precise solar p-mode frequencies are often determined
from time series that contain many years of data. One drawback to
examining such long data sets is that, because of the 11-yr solar
cycle, they must span a wide range of different activity levels.
Since solar p-mode frequencies, lifetimes, powers, and peak
asymmetries all vary systematically with solar activity \citep[][and
references therein]{Howe2008} care must be taken when analysing the
mode parameters obtained from long sets of data. For example,
between solar maximum and minimum the frequency of a low-$l$ mode at
approximately $3000\,\rm \mu Hz$ can change by as much as $1\,\rm\mu
Hz$, an amount that is far greater than the errors associated with
frequency estimates.

\citet{Broomhall2009} has recently published a list of definitive
low-$l$ frequencies that were obtained from $8640\,\rm d$ of BiSON
data. The authors list the raw frequencies, which were determined
directly from the data by fitting profiles to the frequency-power
spectrum. \citeauthor{Broomhall2009} also quote frequencies that
have been corrected for various solar cycle effects. This paper
describes, in detail, how these corrections were made in a manner
that will allow the reader to reproduce the corrections with their
own data. This will allow comparisons between the frequencies
obtained from data sets that were observed at different epochs and
consequently different activity levels.

We begin, in Sect. \ref{section[linear]}, by describing the
well-known linear solar cycle correction that can be used to correct
mode frequencies to a nominal activity level. Then, in Sect.
\ref{section[devil]}, we describe how to determine the
`devil-in-the-detail' correction that accounts for cross-talk
between the variations of different mode parameters with the solar
cycle, and the distribution of activity levels over the period of
observations. Finally, Sect. \ref{section[sas]} describes a
correction that is necessary when using Sun-as-a-star observations
because not all mode components are detectable.

\section{Linear solar cycle correction}\label{section[linear]}
The raw frequencies determined for two sets of data will be
different if they were observed at different epochs. To allow
frequencies from different periods of time to be compared, a solar
cycle correction must be performed. This correction also allows the
frequencies that would have been observed at the canonical quiet Sun
level to be determined. The correction is based on the assumption
that variations in global activity indices can be used as proxies
for low-$l$ frequency shifts. We also assume that the correction can
be parameterised as a linear function of the chosen activity
measure. For the  10.7-cm radio flux \citep{Tapping1990}, these
assumptions are robust \citep{Chaplin2004} at the level of precision
of the data. \citet{Chaplin2007} find that the NOAA Mg II H
core-to-wing ratio \citep{Viereck2001} is predominantly a better
proxy for making solar cycle corrections. However, we found that the
fill of the Mg II data was too sparse to use here.

Let $\nu_{n,l}$ be the set of fitted frequencies determined using
data that were observed during a time when the mean activity level
was $\langle A(t)\rangle$. For example, the mean 10.7-cm radio flux
observed in the 8640-d time series examined by \citet{Broomhall2009}
was $\langle A(t)\rangle=118\times10^{-22}\,\rm W\,m^{-2}\,Hz^{-1}$.
(For the remainder of this paper we refer to $10^{-22}\,\rm
W\,m^{-2}\,Hz^{-1}$ as radio flux units, RFU.) This value of
$\langle A(t)\rangle$ has been calculated taking into account any
gaps that were present in the BiSON time series (e.g. due to
inclement weather and very occasionally instrumental problems). Let
$\langle A(t) \rangle_c$ be the activity level we wish to correct
to. For example, \citeauthor{Broomhall2009} quote frequencies that
have been corrected to the canonical quiet-Sun value of the radio
flux, which is fixed, from historical observations, at $\langle
A(t)\rangle_c=64\,\rm RFU$ \citep{Tapping1990}. The linear solar
cycle correction, $\delta\nu_{n,l}$, is given by

\begin{equation}\label{equation[linear correction]}
\delta\nu_{n,l}=-g_l\mathcal{F}(\nu)\left[\langle
A(t)\rangle-\langle A(t)\rangle_c\right],
\end{equation}
where $g_l$ are $l$-dependent factors that calibrate the size of the
shift and were determined in the manner described below. The
$\mathcal{F}(\nu)$ in equation \ref{equation[linear correction]} is
a function that allows for the dependence of the shift on mode
frequency. We have used the same frequency dependence as was
determined by \citet{Chaplin2004a,Chaplin2004}. The calculated
$\delta\nu_{n,l}$ can be added to the set of frequencies,
$\nu_{n,l}$, to obtain frequencies that are corrected to a mean
activity level of $\langle A(t) \rangle_c$.

\begin{table}
\centering \caption{Values of $g_l$ used to make the linear solar
cycle corrections in \citet{Broomhall2009}.}\label{table[g values]}
\begin{tabular}{cc}
  \hline
  $l$ & $\quad g_l\,\,(\times10^{-3}\,\rm \mu Hz\,RFU^{-1})\quad$ \\
  \hline
  0 & $2.3\pm0.1$ \\
  1 & $3.1\pm0.1$ \\
  2 & $3.1\pm0.1$ \\
  2 & $3.3\pm0.2$ \\
  \hline
\end{tabular}
\end{table}

\begin{figure}
  \centering
  \includegraphics[height=4cm, clip]{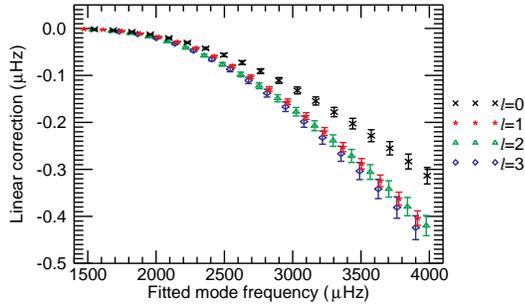}\\
  \caption{Linear corrections that were applied to the
  frequencies observed in the 8640-d time series, where the
  average activity level was $118\,\rm RFU$, to obtain frequencies correct for the canonical quiet-Sun level of
  $64\,\rm RFU$.}\label{figure[linear]}
\end{figure}

To determine $g_l$ the 8640-d set of BiSON data was split into 20
contiguous $432\,\rm d$ segments. To uncover the dependence of the
solar-cycle frequency shifts on the 10.7-cm radio flux mode
frequencies were determined for each of these 20 time series in the
manner described by \citet{Chaplin2004}. The gradient of the linear
relationship between activity and frequency varies significantly
with $l$ because of the spatial dependence of the surface activity.
Table \ref{table[g values]} gives the values of $g_l$ used in
\citet{Broomhall2009}. Figure \ref{figure[linear]} shows the linear
frequency shifts, $\delta\nu_{n,l}$, that were applied to the 8640-d
BiSON data to correct the frequencies to a mean activity level of
$64\,\rm RFU$.

The errors on $\delta\nu_{n,l}$
are dominated by the uncertainties associated with $g_l$. These
errors need to be propagated through to the corrected frequencies
and so the corrected uncertainties are of the order of 10 per cent
larger than those associated with the raw fitted frequencies,
$\nu_{n,l}$.

It is possible that the true relationship between the observed
frequency shifts and solar activity is quadratic or even of higher
order. However, we do not find a higher-order relationship when we
analyse the BiSON frequency shifts i.e., a linear relationship
provides the best fit, given the precision in the BiSON frequency
shifts. In order to ensure our correction procedure is internally
consistent, we use BiSON shifts and a linear calibration.

An alternative approach is to use separate corrections for the
rising and falling phases of the cycle. \citet{Chaplin2004} show
that there is a significant difference in the slope of the linear
relationship between the rising and falling phases of the solar
cycle. More recently, \citet{Jain2009} found, when studying modes of
intermediate degree, that different slopes are required to
accurately describe the observed changes in frequency when the cycle
is split into different phases. Such an approach would be worthwhile
if the data for which the solar cycle corrections are being made
span only a rising or a falling phase.

However, the time series considered by \citet{Broomhall2009} spans
8640\,d, including two rising phases and two falling phases. To test
whether it was advantageous to fit the phases separately we have
split the $432\,\rm d$ time series into two sets, one for the rising
phases and one for the falling phases. A linear fit between
frequency shift and activity was performed for each set. A weighted
mean of the linear corrections for the two phases was then
calculated, where the weights corresponded to the proportion of the
8640\,d spent in each phase. We found that for theoretical modes at
$3000\,\rm\mu Hz$ with $l$ in the range $0-3$ the two-phase
frequency correction was in good agreement (within $1\sigma$) of the
corrections found when a single linear fit was applied. Furthermore,
the error bars associated with the two different correction
approaches were similar in size. In other words, when considering a
long time series that spans both rising and falling phases of the
cycle it is more than adequate to use a single linear fit because
the differences between the rising and falling phases are averaged.

\section{Devil-in-the-detail and activity distribution
correction}\label{section[devil]}

\subsection{A description of the devil-in-the-detail effect}

The linear solar cycle correction assumes that the observed mode
frequencies correspond to an unweighted average of the time-varying
frequencies. However, \citet{Chaplin2008} found that when analysing
long data sets this is not the case. Long time series span, at the
very least, a sizeable fraction of the 11-yr solar activity cycle.
The data examined by \citet{Broomhall2009} is a good example of such
a time series as it spans more than two complete solar cycles.
\citet{Chaplin2008} find that the frequencies observed in long time
series are biased by a cross-talk effect that has its origins in the
simultaneous variations of mode frequencies, powers, and linewidths
over the solar cycle. \citeauthor{Chaplin2008} call this bias to the
observed frequencies the `devil-in-the-detail' effect and we now
describe in detail why this effect is observed.

Consider a simplistic scenario where only the mode frequencies
change with time. The observed frequency of a mode is then
determined by the activity levels observed over the total period of
observations and the unshifted frequency of the mode in question.
Now consider the fictitious scenario where we have 100\,d of data
and on 99 of the days the solar activity level is constant. Then on
the $100\rm^{th}$ day the activity level doubles.

The final observed profile of a mode is well represented by the sum
of the mode profiles observed on each of the 100\,d, normalised by
the length of the observations. For the first 99\,d of our
fictitious time series the same profile is observed. However, on the
last day the peak frequency of the mode profile will be shifted by
an amount, given by equation \ref{equation[linear correction]},
which is dependent on the mode's degree and frequency. The influence
of the final day on the peak frequency of the integrated mode
profile will depend on the size of the frequency shift experienced
by the mode and the mode's linewidth. If the frequency shift is
significantly smaller than the mode linewidth the final-day profile
will be approximately equal to the unweighted average of the daily
mode frequencies. However, if the frequency shift of the mode is
large compared to its linewidth the peak of the profile for the
final day will be situated in the wing of the integrated mode
profile. Consequently, the integrated profile's peak frequency will
not equal the unweighted average of the daily mode frequencies.
Effectively, the final peak frequency of the integrated profile is a
weighted average of the time-varying frequencies, where the weights
are a function of the frequency shift observed on each day and the
width of the mode profile.

\begin{figure}
  \centering
  \includegraphics[height=4cm, clip]{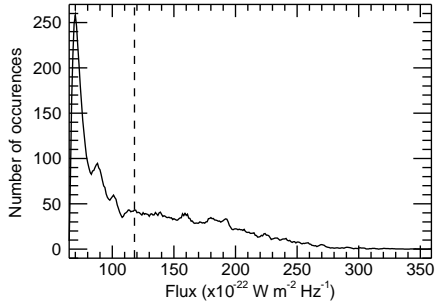}\\
  \caption{Distribution of the 10.7-cm radio flux over
  the 8640-d time series examined by \citet{Broomhall2009}.
  The solid line shows the flux distribution and the dashed line
  indicates the mean flux of the complete time series once the BiSON
  window function has been accounted for (i.e. $118\,\rm RFU$).}
  \label{figure[activity distribution]}
\end{figure}

The above example is extreme and the described effect is not
important if the activity levels observed during the period of
observations are uniformly distributed. However, as can be seen from
Fig. \ref{figure[activity distribution]} the activity distribution
over the 8640-d observations examined by \citet{Broomhall2009} is
heavily skewed towards low activity levels. At low activity levels
the observed frequencies are at a minimum. Therefore the frequency
of the integrated profile will be smaller than that expected from an
unweighted average of the instantaneous frequencies.

This is not the only effect that influences the final observed
frequency. As mentioned, it is simplistic to assume that only the
mode frequencies change with activity. In fact the heights and
widths of the mode profiles also vary throughout the solar cycle and
this means that the data observed at different epochs have varying
contributions to the final peak profile. For typical low-$l$ modes,
as the surface activity increases the mode frequencies and widths
increase but the mode heights decrease. Data samples from times when
the mode is more prominent (i.e. at low activity levels) carry
proportionately larger weights in the final peak profile than times
when the mode is less prominent (i.e. at times of high activity).
Therefore, data samples from times when the frequencies are lower
carry proportionately larger weights in determining the final peak
profile and so the final observed frequency is biased towards a
lower frequency than would be observed in an unweighted mean. The
size of the effect is dependent on the comparative time variations
in frequency and power.

We now describe how the magnitudes of both of these effects were
determined for the 8640-d time series examined by
\citet{Broomhall2009}

\subsection{How the devil-in-the-detail correction was
determined}\label{devil method}

\citet{Chaplin2008} find that the final mode profile observed in a
long data set is well described by the integral over time of the
instantaneous Lorentzian profiles sampled at time, $t$, which have
the appropriate frequency, power, and damping rate for that time and
activity level, and so
\begin{equation}\label{equation[profile integration]}
    \langle
    P(\nu)\rangle=\frac{1}{T}\int_{t=0}^T\frac{H(t)}{1+\xi(t)}\textrm{d}t,
\end{equation}
where
\begin{equation}\label{equation[xi]}
    \xi(t)=\frac{2\left[\nu-\nu(t)\right]}{\Delta(t)}.
\end{equation}
Here, $H(t)$ is the instantaneous peak height of the profile, $T$ is
the total observing time, $\nu(t)$ is the frequency of
the mode at time, $t$, and $\Delta(t)$ is the instantaneous width of the mode
profile.

To determine the size of the frequency bias that occurs because of
the distribution of activity levels we consider the simplistic case
of a single mode whose peak height, $H(t)$, and width, $\Delta(t)$,
remain constant with time. However, we take the peak frequency of
the profile, $\nu(t)$, to be a linear function of the activity at
time, $t$.

\begin{figure}
\centering
  \includegraphics[height=4cm, clip]{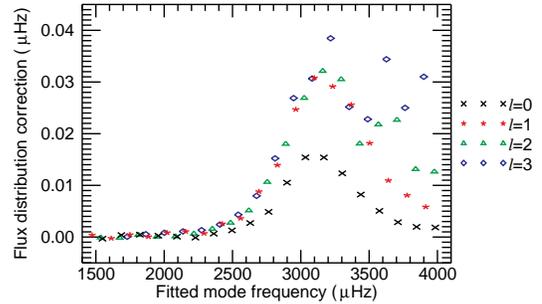}\\
  \caption{Corrections applied to the 8640-d frequencies
  that occur due to the skewed distribution of the observed
  activity level i.e. the devil-in-the-detail corrections calculated
  when the mode heights and widths are kept constant with time.}\label{figure[devil const h and w]}
\end{figure}

For each day in the 8640-d time series we have produced a mode
profile whose frequency was determined by the 10.7-cm radio flux
observed on that day i.e. we use equation \ref{equation[linear
correction]} to determine the frequency of the mode at time, $t$. We
then summed the individual day Lorentzians to produce a single
profile that was representative of the profile observed over a long
time series. We found that the frequency of the final profile was
not equal to the unweighted average of its time-varying frequency.
Figure \ref{figure[devil const h and w]} shows the frequency
correction applied to correct the observed frequency to the
unweighted average frequency when $H(t)$ and $\Delta(t)$ were
constant. As can be seen the correction is significant and
structured.

\begin{figure}
\centering
  \includegraphics[height=4cm, clip]{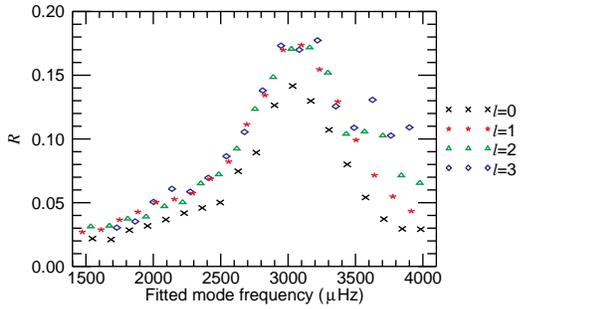}\\
  \caption{Variation in $R$ with mode frequency.}\label{figure[ratio]}
\end{figure}

To understand this effect in more detail we consider the following
ratio
\begin{equation}\label{equation[ratio]}
    R=\frac{\delta\nu_{n,l}}{\Delta_{n,l}},
\end{equation}
where $\delta\nu_{n,l}$ was determined using equation
\ref{equation[linear correction]} for $\langle A(t)\rangle-\langle
A(t)\rangle_c=(118-64)\,\rm RFU=54\,\rm RFU$. Figure
\ref{figure[ratio]} shows $R$ plotted as a function of the mode
frequency. Comparison with Fig. \ref{figure[devil const h and w]}
shows that $R$ has a similar structure to the observed frequency
shift. Hence, the larger the shift experienced relative to the width
of a mode the greater is the difference in the integrated profile
frequency and the unweighted-mean frequency.

In reality the above example is simplistic as the heights and widths
of the mode profiles also vary with the solar cycle. To determine
the size of the frequency shift due to both the distribution of the
activity and the cross-talk between the cycle varying mode
parameters we systematically altered the mode profile widths and
heights with activity.

\begin{figure}
  \centering
  \includegraphics[width=0.24\textwidth, clip]{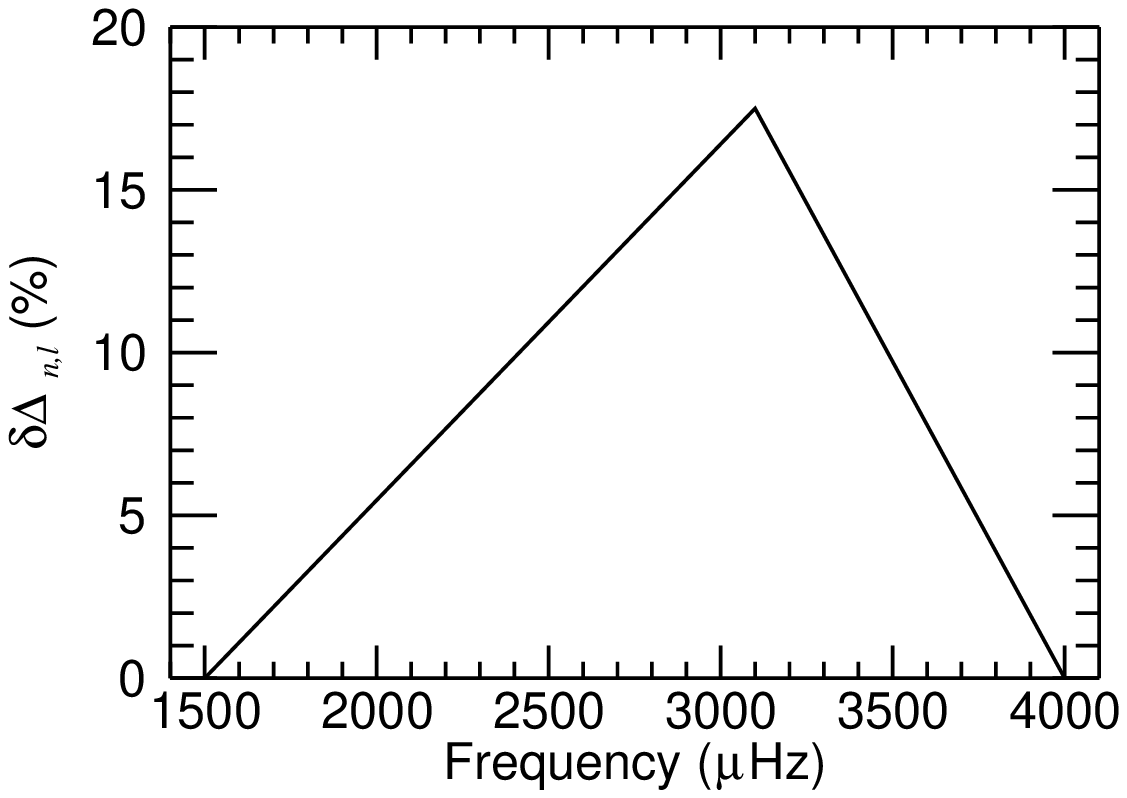}
  \includegraphics[width=0.24\textwidth, clip]{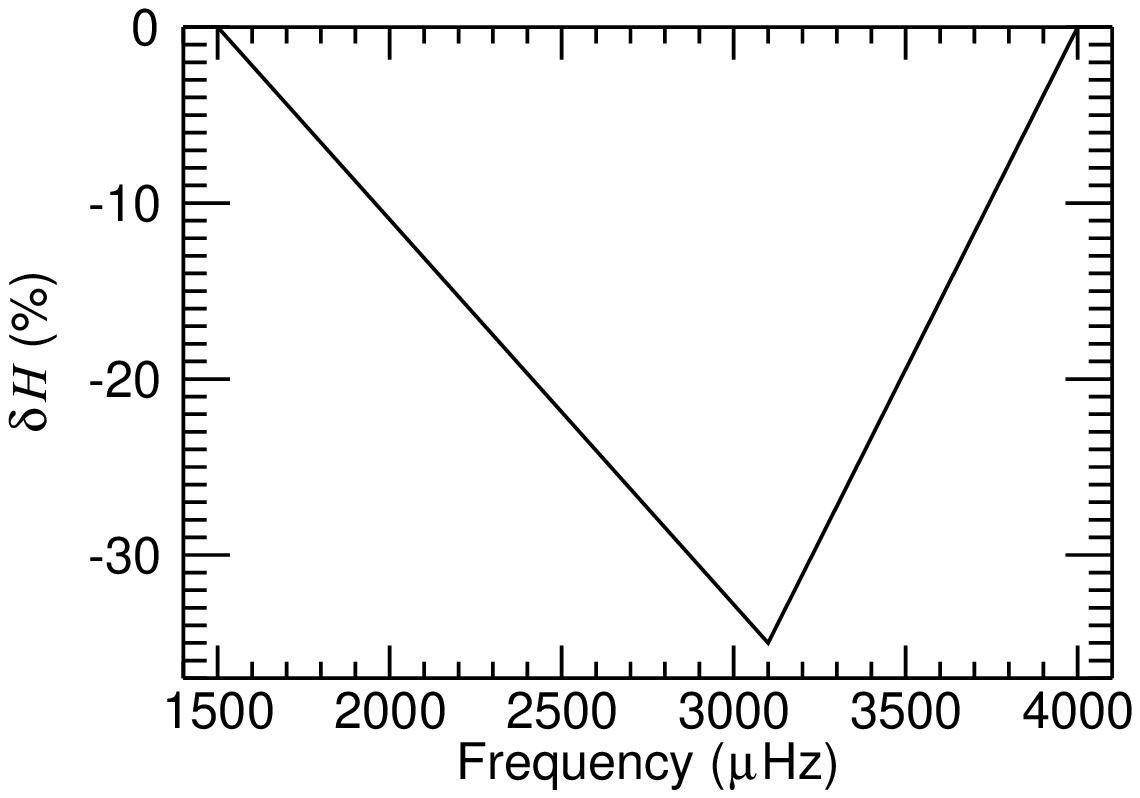}\\
  \caption{Left panel: Model of how the percentage change in the mode
  profile width, $\Delta_{n,l}$, changes with frequency.
  Right panel: Model of how the percentage change in the mode
  profile height, $H$, changes with frequency. Both panels represent the
  maximum change expected and so are applicable when the activity
  level is at a maximum.}\label{figure[profile change]}
\end{figure}

Between solar minimum and maximum mode widths are observed to
increase by up to 17.5 per cent and mode heights are observed to
decrease by twice this amount \citep{Chaplin2000, Salabert2007}. The
variations in mode widths and heights are also functions of
frequency. We have modelled the change in profile widths and heights
in the manner shown in Fig. \ref{figure[profile change]}. Note that
the size of the variation also depends linearly on the level of
activity and the profiles shown in Fig. \ref{figure[profile change]}
represent the maximum changes in width and height respectively,
which occur when the solar activity is at a maximum. We have varied
the frequency of the mode in the same manner as before. We have then
produced instantaneous profiles of a mode for each day in the 8640-d
time series and we have used these profiles to produce an integrated
profile which represents that observed in a long time series.

\begin{figure}
  \centering
  \includegraphics[height=4cm, clip]{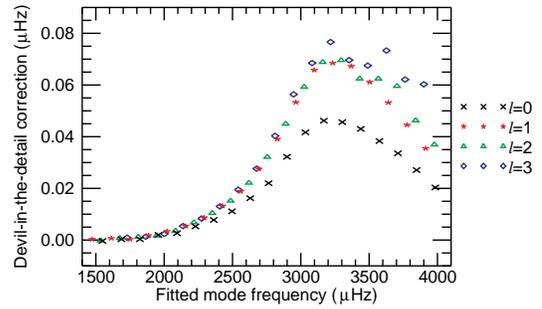}\\
  \caption{Devil-in-the-detail corrections that were used in
  \citet{Broomhall2009}.}\label{figure[devil]}
\end{figure}

The total devil-in-the-detail corrections applied to the 8640-d
BiSON data are plotted in Fig. \ref{figure[devil]}. These
corrections include both the effect of the distribution of the
activity level and the cross-talk between the variations in mode
frequency, width, and height over the observation period. Comparison
with Fig. \ref{figure[devil const h and w]} implies that the
distribution of the activity level accounts for approximately half
of the total devil-in-the-detail correction. The frequency
corrections shown in Fig. \ref{figure[devil]} should be added to the
observed frequencies, $\nu_{n,l}$, to obtain frequencies that have
been corrected for the devil-in-the-detail effect. For the 8640-d
data set considered here the maximum devil-in-the-detail corrections
are of the order of 5 times larger than the errors associated with
the raw fitted frequencies.

\section{Sun-as-a-star correction}\label{section[sas]}
Sun-as-a-star observations are taken from a perspective where the
plane of the Sun's rotation axis is nearly perpendicular to the
line-of-sight. Therefore, only modes where $l+m$ is even have a
non-negligible visibility in Sun-as-a-star observations.
Furthermore, estimates of the centroid frequencies are dominated by
the $|m|=l$ components, which are the most prominent components
observed in the multiplets. It can, therefore, be difficult to
estimate the true centroid frequency of a mode.

The difference between the centroid and fitted frequencies depends
on the level of activity over the period the observations were made.
When the solar activity is at a minimum the components are observed
to be in a near-symmetrical arrangement and so the fitted centroid
frequency is close to the true centroid frequency. However, at
moderate to high activity levels this is not the case as the mode
components are not arranged symmetrically. Take, for example, an
$l=2$ mode. The $|m|=2$ components will experience a larger shift at
high activity levels than the $m=0$ component. The magnitude of
observed asymmetry is related to the inhomogeneous distribution of
the solar activity over the surface and the spherical harmonic
associated with each visible $m$ component. Therefore, the fitted
frequencies differ from the true centroids by an amount that is
dependent on $l$.

\citet{Appourchaux2007} describe how to make a `Sun-as-a-star'
correction, which is determined using the so called $a$ coefficients
that are found from fits for an unresolved Sun. The $a$ coefficients
were obtained from MDI data \citep{Schou1998}, which had the same
activity levels as the BiSON data \citep{Chaplin2004b}. It is
important to make this correction as the centroid frequency contains
information on the spherically symmetric component of the internal
structure and so is the required input for hydrostatic structure
inversions. Furthermore, resolved solar observations are able to
offer direct estimates of the centroid frequencies and so, to
compare the frequencies determined using resolved and Sun-as-a-star
data, the Sun-as-a-star correction should be employed.

\begin{figure}
  \centering
  \includegraphics[height=4cm, clip]{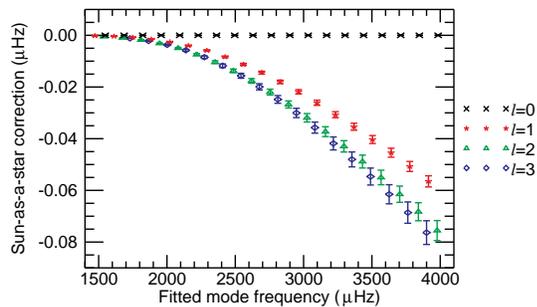}\\
  \caption{Sun-as-a-star corrections that were used in
  \citet{Broomhall2009}.}\label{figure[sas]}
\end{figure}

Figure \ref{figure[sas]} shows the Sun-as-a-star corrections applied
to the 8640-d BiSON data set. As can be seen, except for $l=0$ modes
these corrections are similar in magnitude but opposite in sign to
the devil-in-the-detail corrections. As the variation with frequency
is different for the two different corrections it is still necessary
to use both when examining the frequencies observed in long time
series. The correction to the $l=0$ modes is zero as there is only
one component to observe and by definition this is the centroid
frequency.

\section{Summary}
All three of the corrections described in this paper are dependent
on the solar activity level over the period of the observations and
so should be used when comparing the frequencies determined from
data observed at different epochs. The linear solar cycle correction
allows frequencies to be determined for a nominal activity level.
The devil-in-the-detail correction accounts for the distribution of
activity levels over the period of observations and the fact that
mode profile heights, widths, and frequencies all vary with solar
activity. This correction should be used when comparing frequencies
obtained from data sets of different lengths, although the
correction is only significant when the data sets are long. The
Sun-as-a-star correction allows for the fact that only components
where $l+m$ is even can be detected in Sun-as-a-star data and should
be used when comparing the frequencies obtained from resolved and
Sun-as-a-star data. This is because, at times of high-activity, the
$m$-components of a mode are not arranged symmetrically, and so,
because not all components are visible, Sun-as-a-star observations
measure a frequency that is slightly different to the centroid
frequency.

\section*{Acknowledgements}

We thank the anonymous referee for insightful comments. This paper
utilises data collected by the Birmingham Solar-Oscillations Network
(BiSON). We thank the members of the BiSON team, both past and
present, for their technical and analytical support. We also thank
P. Whitelock and P. Fourie at SAAO, the Carnegie Institution of
Washington, the Australia Telescope National Facility (CSIRO), E.J.
Rhodes (Mt. Wilson, Californa) and members (past and present) of the
IAC, Tenerife. BiSON is funded by the Science and Technology
Facilities Council (STFC).

\bibliographystyle{aa}
\bibliography{cycle_corrections_v1}

\end{document}